\documentclass[showpacs,twocolumn,aps]{revtex4}
\usepackage{amssymb}  
\usepackage{amsmath}
\usepackage{graphicx}
\usepackage{lscape}
\usepackage{booktabs}
\begin{document}
\title{Antimatter production in central Au+Au collisions at 
       $\sqrt{s_{\rm{NN}}}$=200 GeV}
\author{Gang Chen$^{1}$, Yu-Liang Yan$^{2,3}$, De-sheng Li$^1$,
Dai-Mei Zhou$^4$, Mei-Juan Wang $^1$, Bao-Guo Dong$^2$, and Ben-Hao Sa
$^{2,4}$} 
\address{
1 School of Mathematics and Physics, China University of Geoscience, Wuhan 430074, China\\
2 China Institute of Atomic Energy, P.O. Box 275(18), Beijing 102413, China\\
3 School of Physics, Institute of Science, Suranaree University of
Technology, Nakhon Ratchasima 30000, Thailand.\\
4 Institute of Particle Physics, Huazhong Normal University, Wuhan
430082,China}

\begin{abstract}
We have used the dynamically constrained phase space coalescence model
to investigate the production of light nuclei (anti-nuclei) based on the 
1.134$\times 10^7$ hadronic final states generated by the PACIAE model for the 
0-5\% most central Au+Au collisions at $\sqrt{s_{\rm{NN}}}$=200 GeV with $|y|
<$1 and $p_T<$5 acceptances. The STAR data of $\overline{_{\overline\Lambda}^
3H}$, ${_{\Lambda}^3 H}$, $^3{\overline{He}}$, and $^3{{He}}$ yields and 
ratios are well reproduced by the corresponding PACIAE results. The transverse 
momentum distribution of $\overline{^3{He}}$ (${^3{He}}$) and $\overline{_
{\overline\Lambda}^3H}$ (${_{\Lambda}^3H}$) is also given. It turned out that 
the transverse momentum distribution of light nuclei is close to that of 
the corresponding anti-nuclei.

\end{abstract}

\maketitle

\section{Introduction}
The nucleus-nucleus collisions at top RHIC energy produce an initial hot and 
dense matter (quark-gluon mater, QGM). It has been interpreted as a strongly 
coupled quark-gluon plasma (sQGP) \cite{brah,phob,star,phen}. This is nearly 
a perfect liquid composed of quarks and gluons but is not a free gas-like 
quark-gluon plasma (fQGP) expected by theorists and experimentalists long time
ago.

The anti-nuclei production is great importance in the nuclear and particle 
physics, the astrophysics, and the cosmology. One believes that the matter 
and antimatter exist in equal abundance during the initial stage of the 
universe. However, a mystery exists: how this symmetry got lost in the 
evolution of the universe with no significant amount of antimatter being 
present. Because the initial fireball created in ultra-relativistic heavy ion 
collisions is similar to the initial stage of the universe, the study of
anti-nuclei production in ultra-relativistic heavy ion collisions may light 
this issue. However, the study of light nuclei (anti-nuclei) production is
quite hard both experimentally and theoretically because of the low 
production multiplicity.

The STAR collaboration has reported their measurements of $^3_{\Lambda}H$ and 
$\overline{_{\overline\Lambda}^3H}$ in Au+Au collisions at the top RHIC energy 
\cite{star2}. They have measured ``70$\pm$17 antihypertritons ($\overline{_
{\overline\Lambda}^3H}$) and 157$\pm30$ hypertritons ($^3_{\Lambda}H$)" in the 
89 million minimum-bias and 22 million central (``head-on") Au+Au collision
events at $\sqrt{s_{\rm{NN}}}$=200 GeV. Thus the corresponding yields are 
estimated to be 6.31$\times 10^{-7}$ and 1.41$\times 10^{-6}$, respectively. 
The ALICE collaboration has also published their preliminary $\overline d$ 
yield of $\sim6\times 10^{-5}$ measured in the pp collisions at $\sqrt s$=7 
TeV \cite{alice,alice1}. 

On the other hand, the theoretical study of light nuclei (anti-nuclei) is
usually separated into two steps. The nucleons and hyperons are first  
calculated with some selected models, such as the transport models. Then the 
light nuclei (anti-nuclei) are calculated by the phase space coalescence
model \cite{grei,chen,ma} and/or the statistical model \cite{pop,peter} etc.
Recently, production of light nuclei (hypernuclei) in Au+Au/Pb+Pb collisions 
at relativistic energies have been investigated theoretically by the 
coalescence+blast-wave method \cite{ma1} and the UrQMD-hydro hybrid model+
thermal model \cite{stei}, respectively.

We have proposed an approach studying the light nuclei (anti-nuclei) 
production in ultra-relativistic pp collisions by dynamically constrained 
phase-space coalescence model \cite{yuyl}. This approach is based on the 
final hadronic state generated by a parton and hadron cascade model PACIAE 
\cite{sa2}. The calculated light nuclei (anti-nuclei) yield in non-single 
diffractive (NSD) pp collisions at $\sqrt{s }$=7 TeV is well comparing with 
ALICE data \cite{alice,alice1}. The transverse momentum distribution and 
rapidity distribution are also predicted for $^3{He}$ ($\overline{^3{He}}$) 
and $_{\overline\Lambda}^3H$ ($\overline{_{\overline\Lambda}^3H}$) in NSD pp 
collisions at $\sqrt{s}$=7 and 14 TeV. In this paper, we use this method 
to investigate the light nuclei (anti-nuclei) and hypernuclei 
(anti-hypernuclei) productions in the Au+Au collisions at $\sqrt{s_{\rm{NN}}}$
=200 GeV.

The paper is organized as follows: In the Sec. II, we briefly introduce the 
PACIAE model and the dynamically constrained coalescence model. In  
Sec. III, the calculated light nuclei (anti-nuclei) and hypernuclei 
(anti-hypernuclei) yields, ratios, as well as the transverse momentum 
distributions are given and compared with the STAR data. A short summary is 
the content of Sec. IV.
\section {MODELS}
The PYTHIA model \cite{sjo2} is devised for the high energy hadron-hadron 
($hh$) collisions. In this model, a hh collision is decomposed into the  
parton-parton collisions. The hard parton-parton scattering is described by 
the leading order perturbative QCD (LO-pQCD) parton-parton interactions with 
the modification of parton distribution function in a hadron. The soft 
parton-parton collision, a non-perturbative phenomenon, is considered 
empirically. The initial- and final-state QCD radiations and the multiparton 
interactions are also taken into account. Therefore, the consequence of a hh 
collision is a partonic multijet state composed of di-quarks (anti-diquarks), 
quarks (antiquarks) and gluons, as well as a few hadronic remnants. This is 
then followed by the string construction and fragmentation. A hadronic final 
state is obtained for a hh collision eventually.

The parton and hadron cascade model PACIAE \cite{sa2} is based on PYTHIA and 
is devised for the nucleus-nucleus collisions mainly. In the PACIAE model, 
first of all the nucleus-nucleus collision is decomposed into the 
nucleon-nucleon (NN) collisions according to the collision geometry and 
NN total cross section. Each NN collision is described by the PYTHIA model 
with the string fragmentation switches-off and the di-quarks (anti-diquarks) 
randomly breaks into quarks (anti-quarks). So the consequence of a NN 
collision is now a partonic initial state composed of quarks, anti-quarks, 
and gluons. Provided all NN collisions are exhausted, one obtains a 
partonic initial state for a nucleus-nucleus collision. This partonic initial 
state is regarded as the quark-gluon matter (QGM) formed in the relativistic 
nucleus-nucleus collisions. Secondary, the parton rescattering proceeds. The 
rescattering among partons in QGM is randomly considered by the 2 
$\rightarrow$ 2 LO-pQCD parton-parton interaction cross sections \cite{comb}. 
In addition, a $K$ factor is introduced here to include the higher order and 
the non-perturbative corrections. Thirdly, the hadronization follows after 
the parton rescattering. The partonic matter can be hadronized by the Lund 
string fragmentation regime \cite{sjo2} and/or the phenomenological 
coalescence model \cite{sa2}. Finally, the hadronic matter proceeds 
rescattering until the hadronic freeze-out (the exhaustion of the 
hadron-hadron collision pairs). We refer to \cite{sa2} for the details.

In quantum statistical mechanics \cite{kubo} one can not precisely 
define both position $\vec q\equiv (x,y,z)$ and momentum $\vec p\equiv 
(p_x,p_y,p_z)$ of a particle in the six dimension phase space, because of 
the uncertainty principle
\begin{equation*}
\Delta\vec q\Delta\vec p\sim h^3.
\end{equation*}
We can only say this particle lies somewhere within a six dimension quantum 
``box" or ``state" with volume of $\Delta\vec q\Delta\vec p$. A particle 
state occupies a volume of $h^3$ in the six dimension phase space \cite{kubo}. 
Therefore one can estimate the yield of a single particle by
\begin{equation}
Y_1=\int_{H\leqslant E} \frac{d\vec qd\vec p}{h^3},
\end{equation}
where $H$ and $E$ are the Hamiltonian and energy of the particle, 
respectively. Similarly, the yield of N particle cluster can be estimated by
\begin{equation}
Y_N=\int ...\int_{H\leqslant E} \frac{d\vec q_1d\vec p_1...d\vec
q_Nd\vec p_N}{h^{3N}}. \label{phas}
\end{equation}

Therefore the $\overline{_{\overline\Lambda}^3H}$ yield in our dynamically 
constrained phase space coalescence model, for instance, is assumed to be
\begin{align}
Y_{\overline{_{\overline\Lambda}^3H}}=&\int ...
\int\delta_{123}\frac{d\vec q_1d\vec p_1
  d\vec q_2d\vec p_2d\vec q_3d\vec p_3}{h^{9}},
\label{yield} \\
\delta_{123}=&\left\{
  \begin{array}{ll}
  1 \hspace{0.2cm} \textrm{if} \hspace{0.2cm} 1\equiv \bar p, 2\equiv \bar n,
    3\equiv \bar\Lambda;\\
    \hspace{0.75cm} m_0\leqslant m_{inv}\leqslant m_0+\Delta m;\\
    \hspace{0.75cm} |\vec q_{12}|\leqslant D_0, \hspace{0.2cm}|\vec q_{13}|
    \leqslant D_0, \hspace{0.2cm}|\vec q_{23}|\leqslant D_0; \\
  0 \hspace{0.2cm}\textrm{otherwise},
  \end{array}
  \right.
\label{yield1}
\end{align}
where
\begin{equation}
m_{inv}=[(E_1+E_2+E_3)^2-(\vec p_1+\vec p_2+\vec p_3)^2]^{1/2},
\label{yield2}
\end{equation}
here ($E_1,E_2,E_3$) and ($\vec p_1,\vec p_2,\vec p_3$) are the
energy and momentum of particles $\bar p,\bar n,\bar\Lambda$, respectively. 
In Eq.~(\ref{yield1}), $m_0$ and $D_0$ stand for, respectively, the rest mass 
and diameter of $\overline{_{\overline\Lambda}^3H}$, $\Delta m$ refers to the 
allowed mass uncertainty, and $|\vec q_{ij}|=|\vec q_{i}-\vec q_{j}|$ is 
the vector distance between particles $i$ and $j$.
\section{Calculations and results}
As the hadron position and momentum distributions from transport model 
simulation are discrete, the integral over continuous distributions in Eq.~
(\ref{yield1}) should be replaced by the sum over discrete distributions. In 
a single event of the final hadronic state obtained from transport model 
simulation, the configuration of $\overline{_{\overline\Lambda}^3H}$ 
($\bar p$+$\bar n$+$\bar\Lambda$) system can be expressed as
\begin{equation}
C_{\bar p\bar n\bar\Lambda}(q_1,q_2,q_3;\vec p_1,\vec p_2,\vec p_3),
\label{conf}
\end{equation}
where the subscripts $1\equiv \bar p$, $2\equiv \bar n$, $3\equiv \bar
\Lambda$, and $q_1$ refers to the distance between $\bar p$ and the
center-of-mass of $\bar p$, $\bar n$, and $\bar\Lambda$ for instance.
Then the third constraint (diameter constraint) in Eq.~(\ref{yield1}) is 
correspondingly replaced by
\begin{equation}
q_1\leqslant R_0, \ \ \ \hspace{0.2cm} q_2\leqslant R_0, \ \ \
\hspace{0.2cm}q_3\leqslant R_0,
\end{equation}
where $R_0$ refers to the radius of $\overline{_{\overline\Lambda}^3H}$.

Each of the above configuration contributes a partial yield of
\begin{equation}
y_{123}=\left\{
  \begin{array}{ll}
  1 \hspace{0.2cm} \textrm{if} \hspace{0.2cm} m_0\leqslant m_{inv}\leqslant 
    m_0+\Delta m,\\
    \hspace{0.4cm} q_1\leqslant R_0, \hspace{0.2cm} q_2\leqslant R_0, 
    \hspace{0.2cm} q_3\leqslant R_0; \\
  0 \hspace{0.2cm}\textrm{otherwise};
  \end{array}
  \right.
\label{yield3}
\end{equation}
to the $\overline{_{\overline\Lambda}^3H}$. So the total yield of 
$\overline{_{\overline\Lambda}^3H}$ in a single event is the sum of the 
above partial yield over the configurations of Eq.~(\ref{conf}) and their  
combinations. An average over events is required at the end.
\begin{table}[htbp]
\caption{Strange particle rapidity density dN/dy at the mid-rapidity 
         ($|y|<$1 for ${\Lambda}$ and $\overline{\Lambda}$, 
         $|y|<$0.75 for $\Xi^-$ and $\overline{\Xi^-}$) in the 0-5\% 
         most central Au+Au central collisions at 
         $\sqrt{s_{\rm{NN}}}$=200 GeV.}
\begin{tabular}{ccc}
\hline
  \hline \cmidrule[0.03pt](l{0.03cm}r{0.03cm}){1-1}
\cmidrule[0.03pt](l{0.03cm}r{0.03cm}){2-2}
\cmidrule[0.03pt](l{0.03cm}r{0.03cm}){3-3}
\ \ \ Particle type & STAR$^a$ &PACIAE \ \ \ \\
\hline
${\Lambda}$ &$16.7\pm0.2\pm1.1$ &16.8 \\
$\overline{\Lambda}$ &12.7$\pm0.2\pm0.9$ &13.1\\
$\Xi^-$ &2.17$\pm0.06\pm0.19$ &2.08\\
$\overline{\Xi^-}$ &1.83$\pm0.05\pm0.20$ &1.68 \\
\hline \hline
\multicolumn{3}{l}{$^a$ The STAR data were taken from \cite{star5}}\\
\end{tabular}
\label{paci1}
\end{table}

In the PACIAE simulations we assume that the hyperons heavier than $\Lambda$ 
already decay. The model parameters are fixed on the default values given in 
PYTHIA. However, the K factor as well as the parameters parj(1), parj(2), and 
parj(3), relevant to the strange production in PYTHIA \cite{sjo2}, are 
roughly fitted to the STAR data of $\Lambda$, $\overline \Lambda$, $\Xi^-$, 
and $\overline\Xi^-$ in the 0-5\% most central Au+Au collisions at $\sqrt{s_
{\rm{NN}}}$=200 GeV \cite{star5}, as shown in Tab.~\ref{paci1}. The fitted 
parameters of $K$=3 (default value is 1 or 1.5 \cite{sjo2}), parj(1)=0.12 
(0.1), parj(2)=0.55 (0.3), and parj(3)=0.65 (0.4) are used to generate 
1.134$\times 10^7$ final hadronic states by the PACIAE model for the 0-5\% 
most central Au+Au collisions at $\sqrt{s_{\rm{NN}}}$ =200 with $|y|<$1 and 
$p_t<$5 acceptances. Then $d$ ($\overline d$), $^3{He}$ ($^3{\overline{He}}$), 
as well as $_{\overline\Lambda}^3H$ ($\overline{_{\overline\Lambda}^3H}$) 
yields and ratios are calculated by the dynamically constrained phase-space 
coalescence model. The calculated results are denoted as ``PACIAE" later. 

Figure~\ref{yield} shows the calculated hadron yields (solid triangles) 
in the Au+Au collisions at $\sqrt{s_{\rm{NN}}}$=200 GeV. The open circles in 
this figure are the experimental data taken from \cite{peter}. One sees in  
this figure that the PACIAE results well agree with the experimental data.

\begin{figure}[htbp]
\includegraphics[width=0.45\textwidth]{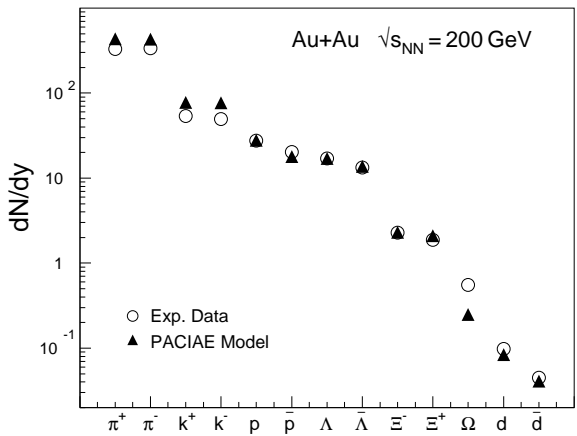}
\caption{Hadron yields (including $d$,$\overline d$) in the Au+Au collisions 
         at $\sqrt{s_{\rm{NN}}}$=200 GeV. The open circles are the 
         experimental data taken from \cite{peter} and the solid triangles 
         are the PACIAE results.} 
\label{yield}
\end{figure}

\begin{table}[htbp]
\caption{Light nuclei (anti-nuclei) rapidity density dN/dy at the midrapidity 
         ($|y|<$1) in the Au+Au collisions at $\sqrt{s_{\rm{NN}}}$=200 GeV.}
\begin{tabular}{ccccc}
\hline \hline 
\cmidrule[0.25pt](l{0.03cm}r{0.03cm}){1-1}
\cmidrule[0.25pt](l{0.03cm}r{0.03cm}){2-2}
\cmidrule[0.25pt](l{0.03cm}r{0.03cm}){3-3}
\cmidrule[0.25pt](l{0.03cm}r{0.03cm}){4-4}
\cmidrule[0.25pt](l{0.03cm}r{0.03cm}){5-5}
Nucleus &Exp. data &PACIAE & Ref.~\cite{ma1}$^a$ &Ref.~\cite{stei}$^b$\\ 
\hline
$ d$ & 0.098$^c$ &0.085$^d$ && \\
$\overline d$ & 0.045$^c$ &0.041$^d$ &&\\
$_{\Lambda} ^3H$ &1.41E-06$^e$& 1.15E-06$^f$ &$>$1.05E-04 &4E-05\\
$\overline{_{\overline\Lambda}^3 H}$ &6.31E-07$^e$&5.29E-07$^f$&$>$4.90E-05&\\
$^3{He}$ &1.72E-06$^g$ &1.50E-06$^f$& $>$1.65E-04&  \\
$^3{\overline{He}}$ &7.09E-07$^h$& 6.17E-07$^f$ & $>$7.30E-05&   \\
\hline \hline
\multicolumn{5}{l}{$^a$ taken from Tab. I in \cite{ma1}.} \\
\multicolumn{5}{l}{$^b$ taken from Fig. 6 in \cite{stei}.} \\
\multicolumn{5}{l}{$^c$ taken from \cite{peter}.}\\
\multicolumn{5}{l}{$^d$ calculated with $\Delta m$=0.0003 GeV.} \\
\multicolumn{5}{l}{$^e$ estimated from \cite{star2}.}\\
\multicolumn{5}{l}{$^f$ calculated with $\Delta m$=0.00015 GeV.} \\
\multicolumn{5}{l}{$^g$ equal to 1.41E-06/0.82 (0.82 is taken from 
                   \cite{star2}).} \\
\multicolumn{5}{l}{$^h$ equal to 6.31E-07/0.89 (0.89 is taken from 
                   \cite{star2}).} \\
\end{tabular}
\label{paci2}
\end{table}
In the Tab.~\ref{paci2} $d$, $\overline d$, $^3He$, $\overline{^3{He}}$, 
$^3_{\Lambda}H$, and $\overline{^3_{\Lambda}H}$ yields are given. The STAR 
yield of $^3{He}$ ($\overline{^3{He}}$) is estimated by the $_{\Lambda} ^3H$ 
($\overline{_{\overline\Lambda}^3 H}$) yield of 1.41E-06 (6.31E-07) divides 
by the ratio of $^3_{\Lambda}H$ to $^3{He}$ 
($\overline{_{\overline\Lambda}^3 H}$ to $\overline{^3_{\Lambda}H}$) of 0.82 
(0.89) \cite{star2}. We see in this table that the agreements between 
experimental data and PACIAE results are well.

We give the ratios of anti-nuclei to nuclei in Tab.~\ref{paci3}. One sees 
again in this table that the PACIAE results are well comparing with the 
STAR data.
\begin{table}[htbp]
\caption{Light anti-nuclei to nuclei ratios in the Au+Au collisions at $\sqrt
         {s_{\rm{NN}}}$=200 GeV. The STAR data are taken from \cite{star2}.}
\begin{tabular}{ccc}
\hline  \hline 
\cmidrule[0.3pt](l{0.03cm}r{0.13cm}){1-1}
\cmidrule[0.25pt](l{0.03cm}r{0.03cm}){2-2}
\cmidrule[0.25pt](l{0.03cm}r{0.03cm}){3-3}
Ratio& STAR &PACIAE \\ 
\hline
$^3{\overline{He}}/^3He$ &0.45$\pm0.18\pm0.07$& 0.41 \\
$\overline{_{\overline\Lambda}^3 H}/_{\Lambda}^3 H$ &0.49$\pm0.18\pm0.07$& 0.46\\
$_{\Lambda}^3 H/^3{He}$ &0.82$\pm0.16\pm0.32$& 0.76 \\
${\overline{_{\overline\Lambda}^3 H}/^3\overline{He}}$ &0.89$\pm0.28\pm0.13$& 0.86\\
\hline \hline
\end{tabular}
\label{paci3}
\end{table}

\begin{figure*}[htb]
\includegraphics[width=0.8\textwidth]{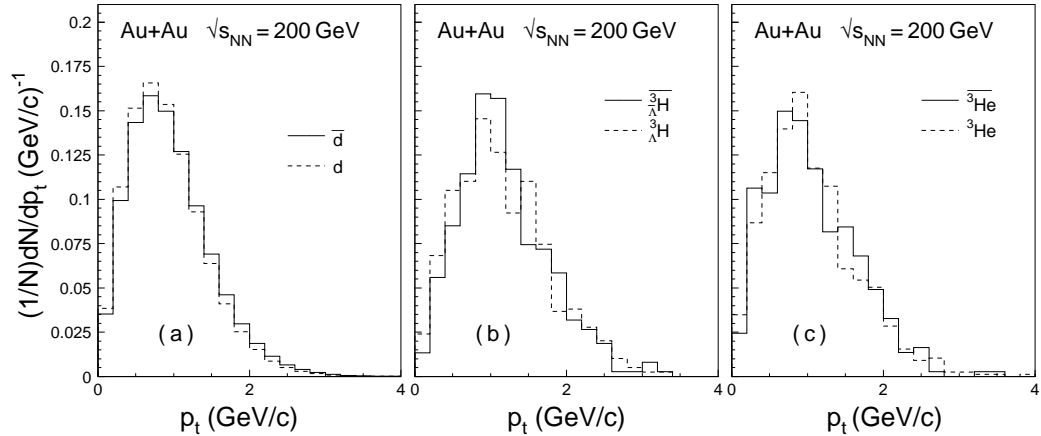}
\caption{(Color online) PACIAE results of the transverse momentum 
distributions of light nuclei (anti-nuclei) in the Au+Au collisions at 
$\sqrt{s_{\rm{NN}}}$=200 GeV. In this figure the blue dashed histograms are 
calculated for nuclei and red solid histograms for anti-nuclei. The panel (a) 
is calculated for $\overline d$ and $d$, (b) for 
$\overline{_{\overline\Lambda}^3 H}$ and ${_{\Lambda}^3 H}$, and (c) 
for $^3{\overline{He}}$ and $^3{{He}}$.} 
\label{tran}
\end{figure*}

We plot in Fig.~\ref{tran} the calculated $d$, $\overline d$, 
${_{\Lambda}^3 H}$, $\overline{_{\overline\Lambda}^3 H}$, $^3{{He}}$, and 
$\overline{^3{He}}$ transverse momentum distributions in the Au+Au collisions 
at $\sqrt{s_{\rm{NN}}}$=200 GeV. Figure ~\ref{tran} (a), (b), and (c) are 
the calculated $p_t$ ditribution for $d$ ($\overline d$), ${_{\Lambda}^3 H}$ 
($\overline{_{\overline\Lambda}^3 H}$), and $^3{{He}}$ ($\overline{^3{He}}$), 
respectively. The pattern of $\overline{^3{He}}$ transverse momentum 
distribution is consistent with the $p_t$ distribution of $\overline{^3{He}}$ 
in the Pb+Pb collisions at $\sqrt{s_{NN}}$=2.76 TeV \cite{alice1}. The peak 
of $p_t$ distribution in the Pb+Pb collision is located at higher $p_t$ than 
the one in the Au+Au collision because of the different reaction energy. 
The strong fluctuation showing in the panel (b) and (c) indicates that the 
1.134$\times10^7$ events are still not enough for these $p_t$ distributions.

\begin{table}[htbp]
\caption{PACIAE results of the light nuclei (anti-nuclei) average 
         transverse momentum $\langle p_t\rangle$ in the Au+Au collisions  
         at $\sqrt{s_{\rm{NN}}}$=200 GeV.}
\begin{tabular}{ccccccc}
\hline \hline
$D$&$\overline D$& $^3{{He}}$& $^3{\overline{He}}$& ${_{\Lambda}^3 H}$&$\overline{_{\overline\Lambda}^3 H}$ &\\
\cmidrule[0.05pt](l{0.9cm}r{0.9cm}){1-7}
\hline
$ 0.92$ \ \ & 0.96 \ \ & 1.05 \ \ & 1.06\ \  &1.15&1.18\ \ \\
\hline \hline
\end{tabular}
\label{paci4}
\end{table}
The PACIAE results of light nuclei (anti-nuclei) average transverse 
momentum in the Au+Au collisions at $\sqrt{s_{\rm{NN}}}$=200 GeV are given in 
Tab.~\ref{paci4}. Here we see that the average transverse momentum of  
light nuclei is nearly equal to the one of corresponding anti-nuclei. This 
feature is already seen in the hadron (anti-hadron) production. 

\section{Conclusion}
In summary, we have employed the dynamically constrained phase space 
coalescence model to investigate the light nuclei (anti-nuclei) production 
based on the final hadronic state generated by the PACIAE model for the 0-5\% 
most central Au+Au collisions at $\sqrt{s_{\rm{NN}}}$=200 GeV with $|y|<$1 
and $p_T<$5 acceptances. The calculated PACIAE results of $d$ ($\overline d$) 
yield of 0.085 (0.041) is close to the experimental datum of 0.098 (0.045) 
\cite{peter}. The PACIAE yields of $\overline{_{\overline\Lambda}^3 H}$, 
${_{\Lambda}^3 H}$, $\overline{^3{He}}$, and $^3{{He}}$ as well as their 
ratios are also consistent with the STAR data \cite{star2}. The PACIAE results 
of light nuclei (anti-nuclei) transverse momentum distributions are also 
given. The consistency between the PACIAE results and the corresponding 
experimental data demonstrates that the PACIAE+dynamically constrained 
phase space coalescence method is able to describe the production of light 
nuclei (anti-nuclei) and hypernuclei (anti-hypernuclei) in the relativity 
heavy ion collisions.
\begin{center} {ACKNOWLEDGMENT} \end{center}
Finally, we acknowledge the financial support from NSFC (11105227,
11075217, 11175070, 11047142, 10975062) in China and SUT-NRU project
(17/2555) in Thailand. GC thanks Dr. Huan Chen for improving the English.

\end{document}